\documentclass[preprint,amsmath,amssymb,aps]{revtex4}
\usepackage{graphicx}
\begin{document}

\title{A simple description of the states $0^+$ and $2^+$ in $^{168}Er$}

\author{A. A. Raduta$^{a),b)}$ and F. D. Aaron $^{a)}$}

\address{$^{a)}$ Department of Theoretical Physics and Mathematics, Bucharest
  University, POBox MG11, Romania}

\address{$^{b)}$ Department of Theoretical Physics, Institute of Physics and
  Nuclear Engineering, Bucharest, POBox MG6, Romania}

\begin{abstract}A sixth-order quadrupole boson Hamiltonian is used to describe 26 states $0^+$ and $67$ states $2^+$ which have been recently identified in $^{168}Er$. Two closed expressions are alternatively used for energy levels. One corresponds to a semi-classical approach while the other one represents the exact eigenvalue of the model Hamiltonian. The semi-classical expression involves four parameters,
while the exact eigenvalue is determined by five parameters. In each of the two descriptions a least square fit procedure is adopted.
 Both expressions provide a surprisingly good agreement with the experimental data.  
\end{abstract}

\maketitle
\renewcommand{\theequation}{\arabic{equation}}
\setcounter{equation}{0}

The collective states of deformed nuclei are usually classified in rotational bands distinguished by a quantum number K, which is the angular momentum projection on the $z$ axis of the intrinsic reference frame. The collective character of the states is diminished by increasing the value of K \cite{Bohr,Grei1,Grei2,Ring}. In Ref.\cite{Rad1} one of us (A.A.R.) suggested a possible method of developing bands in a {\it horizontal} fashion.
Indeed, therein on the top of each state in the ground band a full band of monopole multi-phonon states has been constructed. 
This idea has been recently considered in a phenomenological context trying to organize the states, describing the motion of the intrinsic degrees of freedom, in bands. Thus, two intrinsic collective coordinates, similar to the nuclear deformations $\beta$ and $\gamma$, are described by the irreducible representations of a SU(2) group acting in a fictitious space. Compact formulae for excitation energies have been obtained \cite{Rad2,Rad3}.

In the present letter we address the question whether these expressions can provide a realistic parameterization of the data.
Recently, about 26 states $0^+$ and 67 states $2^+$ have been populated in $^{168}Er$ by means of a $(p,t)$ reaction \cite{Bucur}. In the cited paper the excitation energies and the corresponding reaction strength have been provided. These data were described qualitatively by two microscopic models. One is the projected shell model (PSM) which diagonalizes a microscopic Hamiltonian in a projected basis from a set of intrinsic states,
consisting of zero, two and four quasiparticles. The space of four quasiparticle states excludes the components of four alike quasiparticles. The second model is called quasiparticle phonon model (QPM) and diagonalizes a  microscopic PP+QQ Hamiltonian in a boson space spanned by one and two QRPA phonon states. The final states are restricted to those of low K (=0,1,2). Both models have some inherent drawbacks. PSM restricts the fermion space to four quasiparticle states and even from the four qp space the states with four alike quasiparticles are excluded. This is not the case of QPM where the multi-quasiparticle components are taken into account by means of the QRPA approach. However the final states contain only two phonon states. These states  violate the Pauli principle and moreover are not states of good angular momentum.
Which is the contribution of the spurious components to the calculated observable is not known unless specific additional considerations are made. To conclude, the authors of the mentioned paper interpret the experimental data in terms of single particle degrees of freedom.

By contrast, in the present letter we advance the idea that the microscopic description is not unique and that at least the excitation energies can be realistically parameterized based on a phenomenological description. As we shall show in this letter, the state order number is a quantum number characterizing the intrinsic states.     

In Ref.\cite{Rad3} we have used a sixth-order quadrupole boson Hamiltonian:
\begin{equation}
H=\epsilon \hat{N}+
\sum_{J=0,2,4}C_J\left[\left(b^{\dagger}_2b^{\dagger}_2\right)_J\left(b_2b_2\right)_J\right]_0+F\left(b^{\dagger}_2b^{\dagger}_2\right)_0\hat{N}\left(b_2b_2\right)_0.
\end{equation}
where $b^{\dagger}_{2\mu}, b_{2\mu}$, with $-2\leq \mu \leq 2$, are the quadrupole boson operators and $\hat{N}$ the boson number operator.
This Hamiltonian has been first treated semi-classically. Thus, averaging it on a coherent state for the bosons $b^{\dagger}_{20}$ and $\frac{1}{\sqrt{2}}(b^{\dagger}_{22}+b^{\dagger}_{2,-2})$,
one obtains a classical Hamilton function ${\cal H}$ depending on two coordinates, $q_1$ and $q_2$, and two corresponding conjugate momenta. There are two distinct terms describing an anharmonic motion of a classical plane oscillator and a pseudo-rotation around a certain axis. Taking into account that 
the third component of the pseudo-angular momentum, describing a rotation in a
fictitious space, is a constant of motion the classical Hamiltonian, considered
in the reduced space, can be easily quantized and the resulting energy is:
\begin{equation}
\epsilon_{n,M}=A(n+1)+B(n+1)^2+\frac{C}{2}M^2+\frac{F}{5}\left[(n+1)^3-4(n+1)M^2\right],
\label{eclass}
\end{equation}   
where the factors $A$, $B$ and $C$ have simple expressions in terms of the coefficients
$\epsilon, C_J$ involved in the boson Hamiltonian:
\begin{equation}
A=\epsilon,\; B=\frac{1}{5}C_0+\frac{2}{7\sqrt{5}}C_2+\frac{6}{35}C_4,\;\;C=-\frac{8}{5}C_0+\frac{16}{7\sqrt{5}}C_2-\frac{8}{35}C_4.
\end{equation}
 
Actually, Eq. (\ref{eclass}) represents a semiclassical spectrum which describes the motion of the intrinsic degrees of freedom. Supposing that the rotational degrees of freedom are only weakly coupled to the motion of the intrinsic coordinates, then the total energy associated to the motion in the laboratory frame can be written as a sum of two terms corresponding to the intrinsic and rotational motion, respectively. 
\begin{equation}
\epsilon_{n,M,J}=A(n+1)+B(n+1)^2+\frac{C}{2}M^2+\frac{F}{5}\left[(n+1)^3-4(n+1)M^2\right]+\delta J(J+1)
\label{enmj}
\end{equation}
According to Ref.\cite{Rad3}, to the values $J=0$ and $J=2$ correspond different values of $M$, namely $M=0$ and $M=1$, respectively.
Therefore, considering the above equation for the sets of states with angular momenta $J=0, 2$ and normalizing the results to the energy of the first $0^+$, one obtains the following expressions for the excitation energies:
\begin{eqnarray}
E_{n,0}&\equiv&\epsilon_{n,0,0}-\epsilon_{0,0,0}=
\frac{1}{5}Fn^{3}+(\frac{3}{5}F+B)n^{2}+(A+2B+\frac{3}{5}F)n,\;\; n\geq 0,\nonumber\\
E_{n,1}&\equiv &\epsilon_{n,1,2}-\epsilon_{0,0,0}=  \frac{1}{5}Fn^{3}+(\frac{3}{5}F+B)n^{2}+(A+2B-\frac{1}{5}F)n+{\cal C},\;\; n\geq 1,
\end{eqnarray}
where
\begin{equation}
{\cal C}=-\frac{4}{5}C_0+\frac{2}{7\sqrt{5}}C_2+\frac{6}{35}C_4-\frac{4}{5}F .
\end{equation}
These equations are used to describe the available data for the states $0^+$ and $2^+$ in $^{162}Er$. The coefficients involved in the mentioned equations were fixed by a least square procedure.
The results are:
\begin{equation}
A=394.2\; keV,\;\;
B=-10.4 \;keV,\;\;
F=0.3865 \; keV,\;\;
{\cal C}=-280\;keV.
\end{equation}
\begin{figure}[ht]
\begin{center}
\includegraphics[width=0.5\textwidth]{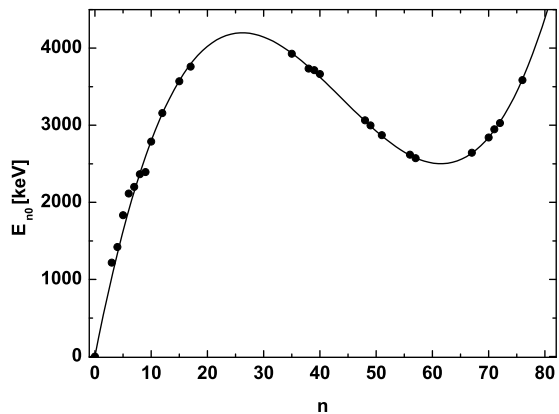}
\end{center}
\caption{The calculated (full curve) excitation energies for the states $0^+$ are compared with the corresponding experimental data. Conventionally, to each experimental data we associate the closest theoretical value $E_{n0}$.}
\label{Fig. 1}
\end{figure}
\begin{figure}[ht]
\begin{center}
\includegraphics[width=0.5\textwidth]{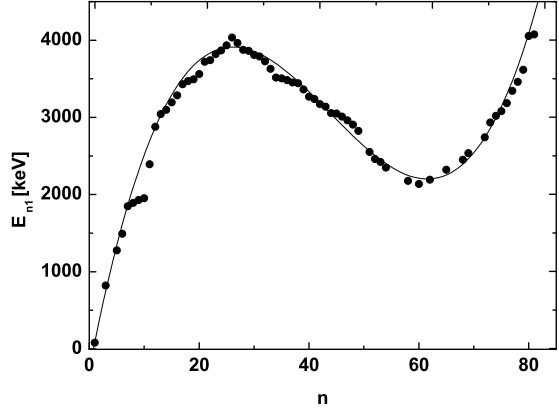}
\end{center}
\caption{The calculated (full curve) excitation energies for the states $2^+$ are compared with the corresponding experimental data. Conventionally, to each experimental data we associate the closest theoretical value $E_{n0}$. }
\label{Fig. 2}
\end{figure}
In order to have a feeling about the fit quality, we compare the results of our calculations with the experimental data in Figs.1 and 2.

Although the starting boson Hamiltonian comprises four and sixth-order boson anharmonicities one can easily derive analytical expressions for its eigenvalues. Indeed, aiming to this goal 
the boson Hamiltonian H is written in an equivalent form:
\begin{equation}
H=(A+\gamma )\hat{N}+(B+\frac{C}{8})\hat{N}^2-\frac{1}{6}\left(B+\frac{C}{8}+\gamma\right)\hat{J}^2-\frac{5}{8}C\left(b^{\dagger}b^{\dagger}\right)_0\left(bb\right)_0.
\end{equation}
where the coefficient $\gamma$ has the expression:
\begin{equation}
\gamma =\frac{2}{7\sqrt{5}}C_2-\frac{3}{7}C_4.
\label{bosh}
\end{equation}
In general, the eigenvalues of a quadrupole boson Hamiltonian are obtained by a diagonalization procedure in the basis $|Nv\alpha JM\rangle$, where the specified quantum numbers are the number of bosons, seniority, missing quantum number, angular momentum and its projection on
the $z$ axis, respectively. 
From Eq. (\ref{bosh}) it is obvious that  $|Nv\alpha JM\rangle$ is an eigenstate of $H$. The corresponding eigenvalues are:
\begin{eqnarray}
E_{N,v,J}&=&\frac{1}{5}FN^{3}+(B+\frac{1}{5}F)N^{2}+ 
(A+\gamma -3(\frac{1}{8}C+\frac{2}{5}F))N-\frac{1}{6}(B+\frac{1}{8}
C+\gamma )J(J+1)\nonumber\\
&+&(\frac{1}{8}C+\frac{2}{5}F)v^{2}+3(\frac{1}{8}C+
\frac{2}{5}F)v
-\frac{1}{5}FNv^{2}-\frac{3}{5}FNv. 
\label{envj}
\end{eqnarray}
Comparing this with Eq.(\ref{enmj}), we notice that the eigenvalues of $H$ are characterized by two quantum numbers, namely the number of bosons $N$ and the seniority $v$.
Therefore, using the new expression for energies one expects a better description of the data.
For $J = 0$ we  use the lowest two values for seniority quantum number, and obtain: 
\begin{eqnarray}
E_{N,0,0}&=&\frac{1}{5}FN^{3}+(B+\frac{1}{5}F)N^{2}+ 
(A+\gamma -\frac{3}{8}C-\frac{6}{5}F)N,\;N=0,2,4,...
\label{jeq0}\\
E_{N,3,0}&=&\frac{1}{5}FN^{3}+(B+\frac{1}{5}F)N^{2}+ 
(A+\gamma -\frac{3}{8}C-\frac{24}{5}F)N+\frac{9}{4}C+
\frac{36}{5}F, N=3,5,7,....\nonumber
\end{eqnarray}
Similarly, for $J=2$ we consider the lowest two allowed seniorities, and obtain:
\begin{eqnarray}
E_{N,1,2}&=&\frac{1}{5}FN^{3}+(B+\frac{1}{5}F)N^{2}+ 
(A+\gamma -\frac{3}{8}C-2F)N-B-\gamma +\frac{3}{8}C+
\frac{8}{5}F, N=1,3,5... \nonumber\\
E_{N,2,2}&=&\frac{1}{5}FN^{3}+(B+\frac{1}{5}F)N^{2}+ 
(A+\gamma -\frac{3}{8}C-\frac{16}{5}F)N-B-\gamma +\frac{9}{8}
C+4F, N=2,4,6,....\nonumber\\
\label{jeq2}
\end{eqnarray}
Equations (\ref{jeq0}) and (\ref{jeq2}) have been used to get the best fit of  energies for the states $0^+$ and $2^+$, respectively. The values for the parameters $A, B, C, F, \gamma$, provided by the least square procedure are:
\begin{equation}
A=83.2319\; keV,\;\;
B=-10.2454\;keV,\;\;
C=24.00\;keV,\;\;
F=0.3865 keV,\;\;
\gamma =299.8638 keV.
\end{equation}
\begin{figure}[ht]
\begin{center}
\includegraphics[width=0.5\textwidth]{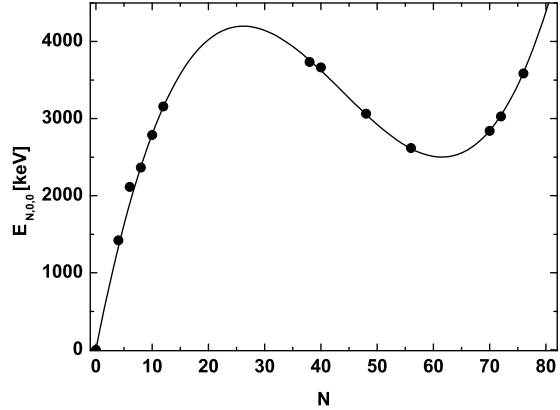}
\end{center}
\caption{The excitation energies calculated with Eq. 11  are compared with the experimental data. }
\label{Fig. 3}
\end{figure}
\begin{figure}[ht]
\begin{center}
\includegraphics[width=0.5\textwidth]{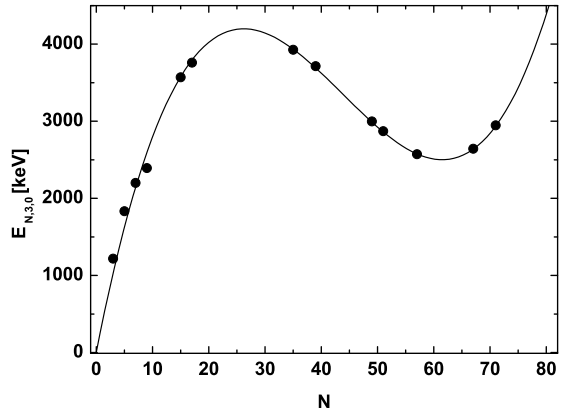}
\end{center}
\caption{The excitation energies characterized by $J^{\pi}=0^+$ and $v=3$ are  compared with the experimental data.}
\label{Fig. 4}
\end{figure}
\begin{figure}[ht]
\begin{center}
\includegraphics[width=0.5\textwidth]{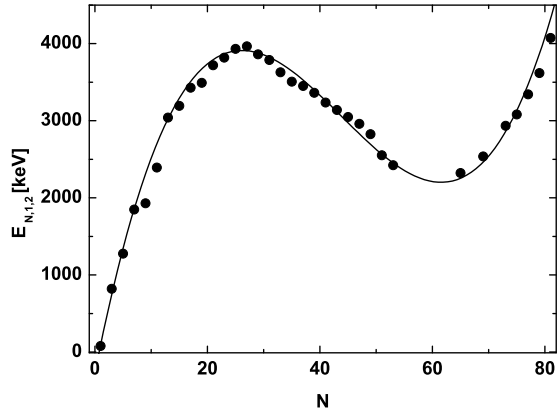}
\end{center}
\caption{Excitation energies with $J^{\pi}=2^+$ and $v=1$ obtained through a least square fit to the experimental data,
represented by black circles.}
\label{Fig. 5}
\end{figure}
\begin{figure}[ht]
\begin{center}
\includegraphics[width=0.5\textwidth]{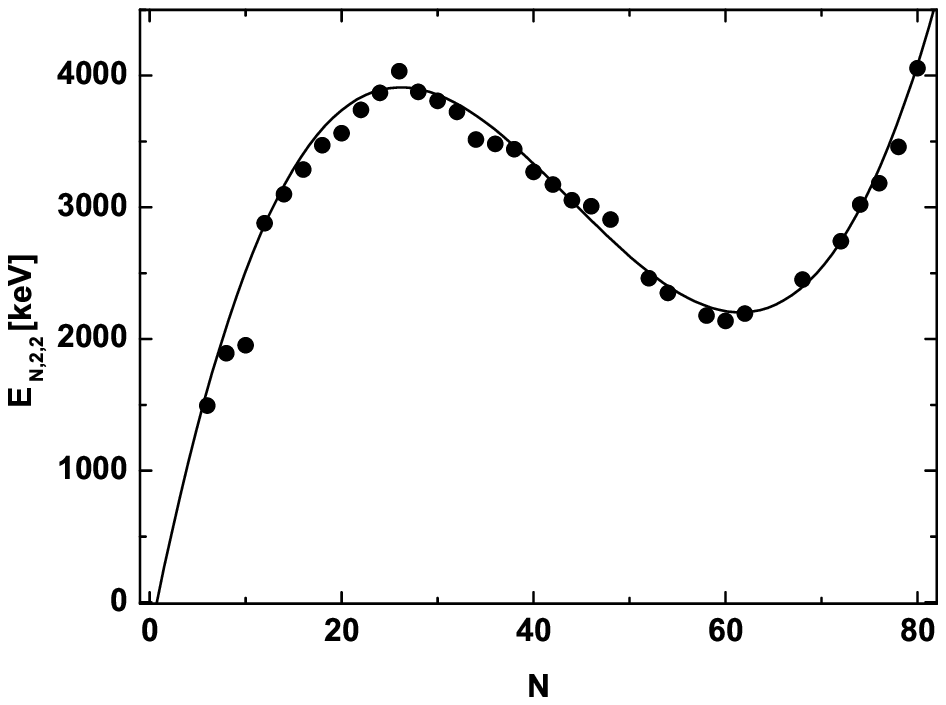}
\end{center}
\caption{Excitation energies of the states characterized by $J^{\pi}=2^+$
and $v=2$ are compared with the experimental data.}
\label{Fig. 6}
\end{figure}
The four sets of energies, calculated with the expressions (\ref{jeq0}) and (\ref{jeq2}), are represented in Figs. 3,4 and 5,6, respectively. The full line curve is the energy as function of N, with N considered as a continuous variable. The  integer number which lies closest to the experimental data is the assigned quantum number. We remark that the agreement with the experimental data is quite good.
The salient feature of our approach is the compact and simple formulae obtained for the excitation energies.
Although in Ref.\cite{Bucur} the energy plot (see Figs. 7,9) has  a logarithmic scale for the ordinate axis, one can see that the agreement shown by Figs. 1-6 of the present letter is better. 

Summarizing, we proposed two  phenomenological descriptions of the excitation energies of the states $0^+$ and $2^+$, seen in $^{168}Er$. 
They correspond to two distinct ways of treating the same sixth-order quadrupole boson Hamiltonian. One is a semiclassical description while the second one uses the exact eigenvalues. While in the yrast band the highest seniority states are the best candidates for a realistic description, for the states of the same angular momentum, conventionally called horizontal bands, the lowest seniority states are used.
It is worth noticing that both $0^+$ and $2^+$ states exhibit a cubic $n$ dependence. We know that such a behaviour for energy in the yrast bands is determining a back-bending \cite{Rege} phenomenon for the moment of inertia as a function of the rotational frequency.
Here a back bending  also shows up but the cause is different from that determining the bending in the moment of inertia in the 
yrast band.  One may argue that for  many of the states, the single particle degrees of freedom prevail. Actually we share this opinion but, on the other hand, we believe that the single particle behaviour may be simulated by the anharmonicities involved.
Certainly,  data  concerning the e.m. transitions of these states are necessary in order to have a complete description.
Usually, whenever the volume of the data to be described is enlarged the formalism which is used is more sophisticated than the one already
applied for a less numerous data.  By contrast, here we use an extremely simple analytical equation, with a relatively small number of 
parameters. Details about the formalism as well as an extensive study of similar data for other nuclei, will be published elsewhere.


\end{document}